\RequirePackage{fix-cm}
\documentclass[smallextended]{svjour3}       
\usepackage{graphicx}
\usepackage{epsfig}
\usepackage{amsfonts}
\usepackage{amsmath}
\usepackage{amssymb}
\usepackage[colorlinks=true,linkcolor=blue,citecolor=magenta,urlcolor=blue]{hyperref}
\newcommand{\notimplies}{\;\not\!\!\!\implies}

\begin{document}


\title{Bell non-locality and Kochen-Specker contextuality: How are they connected?}


\author{Ad\'an Cabello}

\institute{Ad\'an Cabello \at Departamento de F\'{\i}sica Aplicada II and Instituto Carlos~I de F\'{\i}sica Te\'orica y Computacional, Universidad de
	Sevilla, E-41012 Sevilla, Spain \\
	Tel.: +34-954-556671 \\
	Fax: +34-954-556672 \\
	\email{adan@us.es}
}


\date{Received: July 4, 2020/ Accepted: July 3, 2020}

\maketitle


\begin{abstract}
Bell non-locality and Kochen-Specker (KS) contextuality are logically independent concepts, fuel different protocols with quantum vs classical advantage, and have distinct classical simulation costs. A natural question is what are the relations between these concepts, advantages, and costs. To address this question, it is useful to have a map that captures all the connections between Bell non-locality and KS contextuality in quantum theory. The aim of this work is to introduce such a map. After defining the theory-independent notions of Bell non-locality and KS contextuality for ideal measurements, we show that, in quantum theory, due to Neumark's dilation theorem, every quantum Bell non-local behavior can be mapped to a formally identical KS contextual behavior produced in a scenario with identical relations of compatibility but where measurements are ideal and no space-like separation is required. A more difficult problem is identifying connections in the opposite direction. We show that there are ``one-to-one'' and partial connections between KS contextual behaviors and Bell non-local behaviors for some KS scenarios, but not for all of them. However, there is also a method that transforms any KS contextual behavior for quantum systems of dimension $d$ into a Bell non-local behavior between two quantum subsystems each of them of dimension $d$. We collect all these connections in map and list some problems which can benefit from this map. 
\end{abstract}


\section{Introduction}
\label{Sec:1}

\begin{figure}[t!]
	\hspace{4mm}
	\includegraphics[width=10cm]{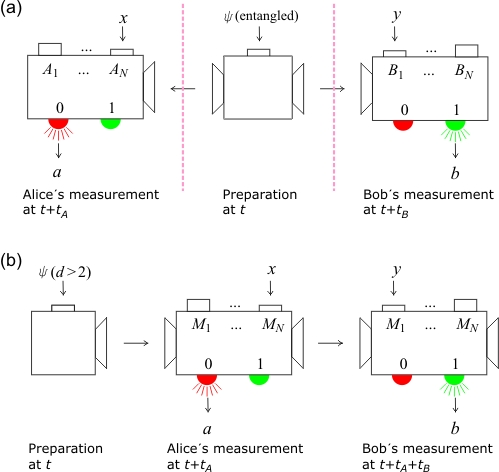}
	\vspace{2mm}
	\caption{(a) Scheme of an experiment of bipartite Bell non-locality on pairs of physical systems. (b) Scheme of an experiment of KS contextuality with sequential ideal measurements on single physical systems.}
	\label{Fig1}
\end{figure}

Bell non-locality \cite{Bell64,CHSH69,BCPSW14} and Kochen-Specker (KS) contextuality \cite{Specker60,Bell66,KS67,KCBS08,Cabello08,Cabello19} are, in principle, logically independent concepts~\cite{BK04}. Both have theory-independent definitions and refer to correlations between outcomes of compatible (or jointly measurable) measurements (i.e., those which can be implemented as different coarse-grains of the same measurement). However, each concept makes emphasis on a different aspect. Bell non-locality focuses on correlations between space-like-separated measurement events on composite systems; see Fig.~\ref{Fig1} (a). KS contextuality does not have the restriction to composite systems or space-like separated events, but it is restricted to ideal \cite{Kleinmann14,CY14,CY16} measurements; see Fig.~\ref{Fig1} (b). Ideal measurements (sometimes called sharp measurements \cite{Kleinmann14,CY14,CY16}) are those that yield the same outcome when they are performed repeatedly on the same physical system and do not disturb the outcome statistics of any compatible observable.
 
In quantum theory, Bell non-locality and KS contextuality fuel different protocols with quantum vs classical advantage. Bell non-locality underpin applications such as device-independent quantum key distribution \cite{Ekert91,BHK05,ABGMPS07}, reduction of communication complexity \cite{BZPZ04}, private randomness expansion \cite{Colbeck06,PAMBMMOHLMM10}, and self-testing \cite{SW87,PR92,MY98}. KS contextuality is a necessary resource for quantum speed-up in some paradigms of fault-tolerant quantum computation \cite{HWVE14,DGBR15,RBDOB17} and is behind the quantum advantage in circuits of bounded depth \cite{BGK18}. 
Simulating with classical systems quantum Bell non-locality requires superluminal communication \cite{Maudlin92,BCT99,TB03}. Simulating KS contextuality with classical systems requires hidden memory \cite{KGPLC11,CGGX18,Budroni19} and has a thermodynamical cost \cite{CGGLW16}.

Nevertheless, in quantum theory, there are many examples where both concepts are deeply connected (see, e.g., \cite{Cabello01,Cabello10,AGA12,CABBB12,LHC16}) and, in fact, it is fair to say that it is quantum KS contextuality what enables quantum Bell non-locality \cite{Cabello19}. However, this view naturally leads to the following questions: 
\begin{itemize}
	\item[(I)] For which pairs of a Bell scenario and a KS scenario, Bell non-locality and KS contextuality are connected ``one-to-one'' in the sense that {\em every} quantum Bell non-local/local behavior (i.e., set of probability distributions, one for each context) violating/non-violating a tight Bell inequality can be mapped to a {\em formally identical} quantum KS contextual behavior violating/non-violating a tight KS non-contextuality inequality which is {\em formally identical} to the precedent Bell inequality?
	\item [(II)] What types of connections can be established between a Bell scenario and a KS scenario when there is no one-to-one connection?
	\item[(III)] What do all these connections tell us about how the respective quantum vs classical advantages and classical simulation costs are related?
\end{itemize}
The aim of this paper is to provide an answer to questions (I) and (II) and list problems for which these answers might be helpful.

The structure of the paper is the following. In Sects.~\ref{Sec:1.1} and \ref{Sec:1.2}, we review the theory-independent definitions of Bell non-locality and KS contextuality for ideal measurements, respectively. In Sect.~\ref{Sec:2}, we show that {\em every} Bell scenario is in one-to-one connection (in the sense defined before) with a KS scenario. Hence, one can write (as, e.g., in \cite{Suarez17}) that
\begin{equation}
\text{\em Bell non-locality} \implies \text{\em KS contextuality}.
\end{equation}
Since KS contextuality also occurs in experiments with indivisible physical systems, it is frequently pointed out (see, e.g., \cite{Suarez17}) that
\begin{equation}
\text{\em KS contextuality} \notimplies \text{\em Bell non-locality}.
\end{equation}
However, in Sect.~\ref{Sec:3.0}, we show that, for certain KS scenarios, every quantum KS contextual behavior can be mapped into a formally identical quantum Bell non-local behavior for a Bell scenario with tight Bell inequalities which are formally identical to the tight KS non-contextuality inequalities of the KS scenario. That is, for some KS scenarios, there is a one-to-one connection in the opposite direction. In addition, in Sect.~\ref{Sec:3.0b} we show that there are other KS scenarios for which there is a connection that is not one-to-one, as some quantum behaviors that are possible in the KS scenario are impossible in the Bell scenario, and a tight KS non-contextuality inequality can map into a Bell inequality that is not tight. In Sect.~\ref{Sec:3.1}, we review a method that converts any quantum KS contextual behavior into a bipartite quantum Bell non-local behavior. Finally, in Sect.~\ref{Sec:6} we collect all the connections in a map and discuss some problems for which this map may be useful.


\subsection{Bell non-locality}
\label{Sec:1.1}


To define Bell non-locality, we need to consider $n\ge 2$ spatially separated observers, here called Alice,$\ldots$, Zoe. We assume that each of them can freely and independently choose one among several possible measurement devices to perform a measurement on its physical system. We will denote by $x,\ldots,z$ Alice's,$\ldots$, Zoe's measurement setting choice, respectively, and by $a,\ldots,c$ their respective outcomes. We also assume that there is space-like separation between any observer's choice and any other observer's outcome. A Bell scenario is characterized by the number of spatially separated observers, the number of measurement devices that each observer has, and the number of possible outcomes that each measurement device has. 

In Bell non-locality, we are interested in theories satisfying the following assumption:

\begin{itemize}
	\item[] {\em Local realism.} The probabilities of outcomes $a,\ldots,c$ for, respectively, measurements $x,\ldots,z$ can be written as
	\begin{equation}
	\label{lr}
	P(a,\ldots,c|x,\ldots,z) = \int d\lambda P(a|x, \lambda) \cdots P(c|z, \lambda),
	\end{equation}
	where $\lambda$ is a set of hidden variables. 
\end{itemize} 
A behavior $\{P(a,\ldots,c|x,\ldots,z)\}$ for a given Bell scenario is Bell non-local if its elements cannot be written as (\ref{lr}).

There are actually two assumptions behind the assumption of local realism. One is that the measurement outcome is associated to local properties. The other is that influences cannot propagate faster than light in vacuum. In Bell non-locality, space-like separation guarantees that any outcome cannot be influenced by the choices made by the distant observers.


\subsection{KS contextuality}
\label{Sec:1.2}


The modern notion of KS contextuality is rooted in the theorems of impossibility of hidden variables in quantum mechanics of Kochen and Specker \cite{Specker60,KS67} and Bell \cite{Bell66}. These theorems are based on two assumptions:
\begin{itemize}
	\item[(I)] Any measurement $x$ that is represented in quantum theory by a self-adjoint operator reveals a preexisting outcome which is determined by the hidden variables and which is the same outcome for all possible sets of commuting measurements that contain $x$ (i.e., sets of measurements represented by mutually commuting self-adjoint operators).
	\item[(II)] The preexisting outcomes of these measurements satisfy the same functional relations that quantum mechanics predicts for quantum systems of a given dimension (for example, for a three-dimensional spin-1 system \cite{KS67,Bell66} or for a four-dimensional pair of spin-1/2 systems \cite{Peres90,Mermin90}). 
\end{itemize} 
In contrast, the modern notion of KS contextuality \cite{Cabello08,Cabello19}, which is the one adopted in this paper, is theory independent. It removes assumption (II) and replaces assumption (I) by:
\begin{itemize}
	\item[] {\em Outcome non-contextuality for ideal measurements.} Any ideal measurement~$x$ reveals a preexisting outcome which is the same for all possible sets of compatible ideal measurements that contain $x$.
\end{itemize}

A KS scenario is characterized by a set of ideal measurements, their respective possible outcomes, and their mutual relations of compatibility. Any set of ideal measurements such that any pair of them is mutually compatible can be jointly measured \cite{CY14}. Therefore, in KS scenarios, contexts are (maximal) sets of compatible ideal measurements.

A behavior $\{P(a,\ldots,c|x,\ldots,z)\}$ for a given KS scenario (i.e., where $x,\ldots, z$ are ideal and compatible measurements) in which each of the measurements has been freely and independently chosen, is KS contextual if it cannot be explained by a model satisfying the assumption of outcome non-contextuality. Hereafter, we will refer to these models as non-contextual hidden-variable (NCHV) models. Equivalently, a behavior produced by ideal measurements is KS contextual if the probability distributions for each context cannot be obtained as the marginals of a global probability distribution on all observables. 

The assumption of outcome non-contextuality for ideal measurements is motivated by the fact that ideal measurements always yield the same outcome no matter which other compatible ideal measurements are performed in between two repetitions of the same measurement. For example, if $x$, $y$, and $z$ are compatible ideal measurements, then performing the sequence $x,y,z,x,z,y,y$ yields the sequence of outcomes $a,b,c,a,c,b,b$, which suggests that measurements reveal persistent properties which can be attributed to the measured system, as different copies of equally designed devices can be used to measure, e.g., $x$ the first and second times.


\section{Every Bell non-local behavior can be mapped to a KS contextual behavior}
\label{Sec:2}


Here, we see in which sense the statement ``Bell non-locality implies KS contextuality'' holds. Consider a Bell inequality for a Bell scenario
\begin{equation}
\label{bi}
\langle \beta \rangle \stackrel{\mbox{\tiny{LR}}}{\leq} \eta,
\end{equation}
where the upper bound $\eta$ holds for theories satisfying local realism (LR). Now consider a Bell experiment aiming a particular quantum Bell non-local behavior that violates this Bell inequality. If all the measurements involved in the experiment would be ideal, then we would be implementing a KS scenario with the same set of measurements, outcomes, and relations of compatibility that the preceding Bell scenario had, and the Bell inequality (\ref{bi}) would become a KS non-contextuality inequality with a formally identical expression, namely,
\begin{equation}
	\label{ni}
	\langle \beta \rangle \stackrel{\mbox{\tiny{NCHV}}}{\leq} \eta,
\end{equation}
where the bound $\eta$, which is the same as in (\ref{bi}), now holds for NCHV theories.
Therefore, a Bell inequality test performed with ideal measurements and yielding a violation of inequality (\ref{bi}) is also a KS contextuality test violating the KS non-contextuality inequality (\ref{ni}). 

What if measurements are not ideal? Then, the experiment is not a KS contextuality test. Nevertheless, in this case, we still can recall a remarkable prediction of quantum theory, namely, that while positive-operator valued measures (POVMs) represent the most general type of measurements allowed in quantum theory, every POVM can be realized as a projection-valued-measure (PVM) in a Hilbert space of augmented dimension. PVMs represent ideal measurements in quantum theory. This prediction follows from Neumark's dilation theorem \cite{Neumark40,Holevo80,Peres95}. This implies that, according to quantum theory, in any Bell scenario, any local measurement represented by a POVM $x$ always admits a local dilation $\hat{x}$ to a local PVM such that $\hat{x}$ is the same in every context of the Bell scenario in which $x$ appears. Therefore, in quantum theory, every Bell non-local behavior $\{P(a,\ldots,c|x,\ldots,z)\}$ can be obtained using only local ideal measurements. ``Generalized'' quantum measurements do not produce Bell non-local behaviors that cannot be produced by ideal measurements. Consequently, according to quantum theory, any Bell experiment producing a behavior $\{P(a,\ldots,c|x,\ldots,z)\}$ can be associated to a KS contextuality experiment which is simply the same Bell experiment but performed with ideal measurements. Therefore, every quantum Bell non-local behavior can be mapped to a formally identical quantum behavior produced by ideal measurements. In particular, every quantum violation of the Bell inequality (\ref{bi}) can be mapped to a quantum violation by the same value of the KS non-contextuality inequality (\ref{ni}).

Moreover, if the Bell inequality (\ref{bi}) is tight, i.e., corresponds to a facet of the polytope of local behaviors \cite{Froissart81,SZ81,Fine82a,Fine82b,Pitowsky89} associated to a Bell scenario with a given set of measurements, outcomes, and relations of compatibility, then the corresponding KS non-contextuality inequality is also tight, as it is associated to the corresponding facet of the corresponding polytope of KS non-contextual behaviors \cite{KBLGC12,AQB13} of the KS scenario with the same measurements, outcomes, and relations of compatibility. 

This link between Bell non-locality to KS contextuality justifies, for instance, considering the experimental tests of the quantum violation of the Clauser-Horne-Shimony-Holt Bell inequality using different degrees of freedom of single photons \cite{MWZ00} and neutrons \cite{YLB03} to be experiments on KS contextuality on indivisible systems. 


\section{Connections from KS contextuality to Bell non-locality}
\label{Sec:3}



\subsection{KS scenarios with complete $n$-partite graphs of compatibility}
\label{Sec:3.0}


There is no general method that converts a quantum KS contextual behavior for an {\em arbitrary} KS scenario into a formally identical quantum Bell non-local behavior for a Bell scenario. However, it is possible to establish a one-to-one connection (as defined before) for some KS scenarios and a less strong connection for other KS scenarios. 

Recall that a KS scenario is characterized by a set of ideal measurements, their outcomes, and their relations of mutual compatibility. These relations can be encoded in a graph in which each measurement is represented by a vertex and only those measurements represented by mutually adjacent vertices are mutually compatible.

In graph theory, an $n$-partite graph is one whose vertices can be divided into $n$ disjoint and independent sets $A$,\ldots, $N$ and such that every edge connects a vertex in one of these sets to one vertex in a different set. Vertex sets $A$,\ldots, $N$ are called the parts of the graph. A $n$-partite graph is complete if there is an edge connecting every vertex of each part with all the vertices of the other parts.

For any $n$-partite Bell scenario the graph of compatibility is complete $n$-partite and has at least two (incompatible) measurements in each part. Therefore, for any KS scenario whose graph of compatibility is complete $n$-partite and has at least two (incompatible) measurements in each part there is a one-to-one connection between any KS contextual behavior and a Bell non-local behavior, and also between tight KS non-contextuality inequalities and tight Bell inequalities.


\subsection{KS scenarios with incomplete $n$-partite graphs of compatibility}
\label{Sec:3.0b}


A more interesting case is that of KS scenarios in which the graph of compatibility is $n$-partite (and has at least two incompatible measurements in each part) but not complete $n$-partite. There, there is a one-to-one connection between any KS contextual behavior and a Bell non-local behavior, and also between tight KS non-contextuality inequalities and Bell inequalities. However, in this case the Bell inequality may not be tight.

The simplest example is the scenario with six dichotomic ideal measurements $M_i$, with $i=1,\ldots,6$, whose graph of compatibility is an hexagon. In this case, the following KS non-contextuality inequality is tight \cite{AQB13}:
\begin{equation}
\label{pearle1}
\langle \gamma \rangle \stackrel{\mbox{\tiny{NCHV}}}{\leq} 4,
\end{equation}
with
\begin{equation}
\label{pearle2}
\langle \gamma \rangle = \sum_{i=1}^5 \langle M_i M_{i+1} \rangle - \langle M_6 M_1 \rangle,
\end{equation}
where $M_i$ has possible results $-1$ and $1$, $M_i$ and $M_{i+1}$ are compatible, and $\langle M_i M_j \rangle$ is the mean value of their products. Here, we can distribute measurements $M_1$, $M_3$, and $M_5$ to Alice (and relabel them as $A_1$, $A_3$, and $A_5$), and measurements $M_2$, $M_4$, and $M_6$ to a spatially separated Bob (and relabel them as $B_2$, $B_4$, and $B_6$). Then,
\begin{equation}
\label{pearle3}
\langle \gamma' \rangle \stackrel{\mbox{\tiny{LR}}}{\leq} 4,
\end{equation}
with
\begin{equation}
\label{pearle4}
\langle \gamma' \rangle = \sum_{i \in \{1,3,5\}} \langle A_i B_{i+1} \rangle - \langle A_1 B_6 \rangle,
\end{equation}
is a Bell inequality. In fact, this is the first ever Bell inequality with more than two alternative settings proposed \cite{Pearle70} (see also \cite{BC90}). Interestingly, it is not a tight Bell inequality. The same happens for the scenarios with an even number $n > 4$ of dichotomic measurements with an $n$-gon as graph of compatibility \cite{AQB13}. 

Therefore, the experiments \cite{BBDH97} can be considered as the first KS contextuality experiments testing a quantum violation of tight KS non-contextuality inequalities different than the KS non-contextuality inequality associated to the bipartite Bell inequality with two measurement settings per party.

The crucial observation is that any quantum violation of a KS non-contextuality inequality in a KS scenario with $n$-partite graph of compatibility (and with at least two incompatible measurements in each part) can be associated to a quantum violation of a formally identical $n$-partite Bell inequality: If the graph of compatibility is complete $n$-partite, then there are one-to-one correspondences between the values of the violations and also between the tightness of the inequalities, but if the graph of compatibility is $n$-partite but not complete, then the quantum value of the violation of the KS non-contextuality inequality can be impossible to achieve in the Bell scenario \cite{XC17} and the tightness of the KS non-contextuality inequality can be lost in the Bell scenario \cite{AQB13}.


\subsection{Connecting arbitrary KS contextuality to Bell non-locality}
\label{Sec:3.1}


Finally, there is a method \cite{Cabello21} which uses the measurements that produce any arbitrary quantum KS contextual behavior in {\em any} KS scenario to produce a (different) quantum Bell non-local behavior in a bipartite Bell scenario. Here we give a general idea of how the method works. For details, see \cite{Cabello21}.

The method begins with the measurements leading to a particular example of state-dependent quantum contextuality. Let us call $S$ this set of measurement. $S$ can be enlarged into a critical state-independent-contextuality (SI-C) set ${\cal S}$. A SI-C set \cite{Cabello08,KBLGC12,BBCP09,YO12,CKB15} is a set of measurements on a quantum system of dimension $d \ge 3$ that produces contextuality for any initial state. Critical means that if we remove any of the elements ${\cal S}$, then the resulting set is not a SI-C set. 

Associated to this SI-C set there is a non-contextuality inequality with bound $\mu$ which is equally violated by all states by a value $q$. 
This inequality can be transformed into a bipartite Bell inequality which has $\mu$ as local bound and, when Alice and Bob share a two-qudit maximally entangled state and choose their measurements in ${\cal S}$, is violated exactly by $q$. The connection between the resulting Bell non-local behavior and the initial KS contextual behavior produced by $S$ comes from the fact that, if we remove from ${\cal S}$ any of the measurements in $S$ then the violation of the Bell inequality (for the maximally entangled state) vanishes. 


\begin{figure}[t!]
	\hspace{-12mm}
	\includegraphics[width=14.2cm]{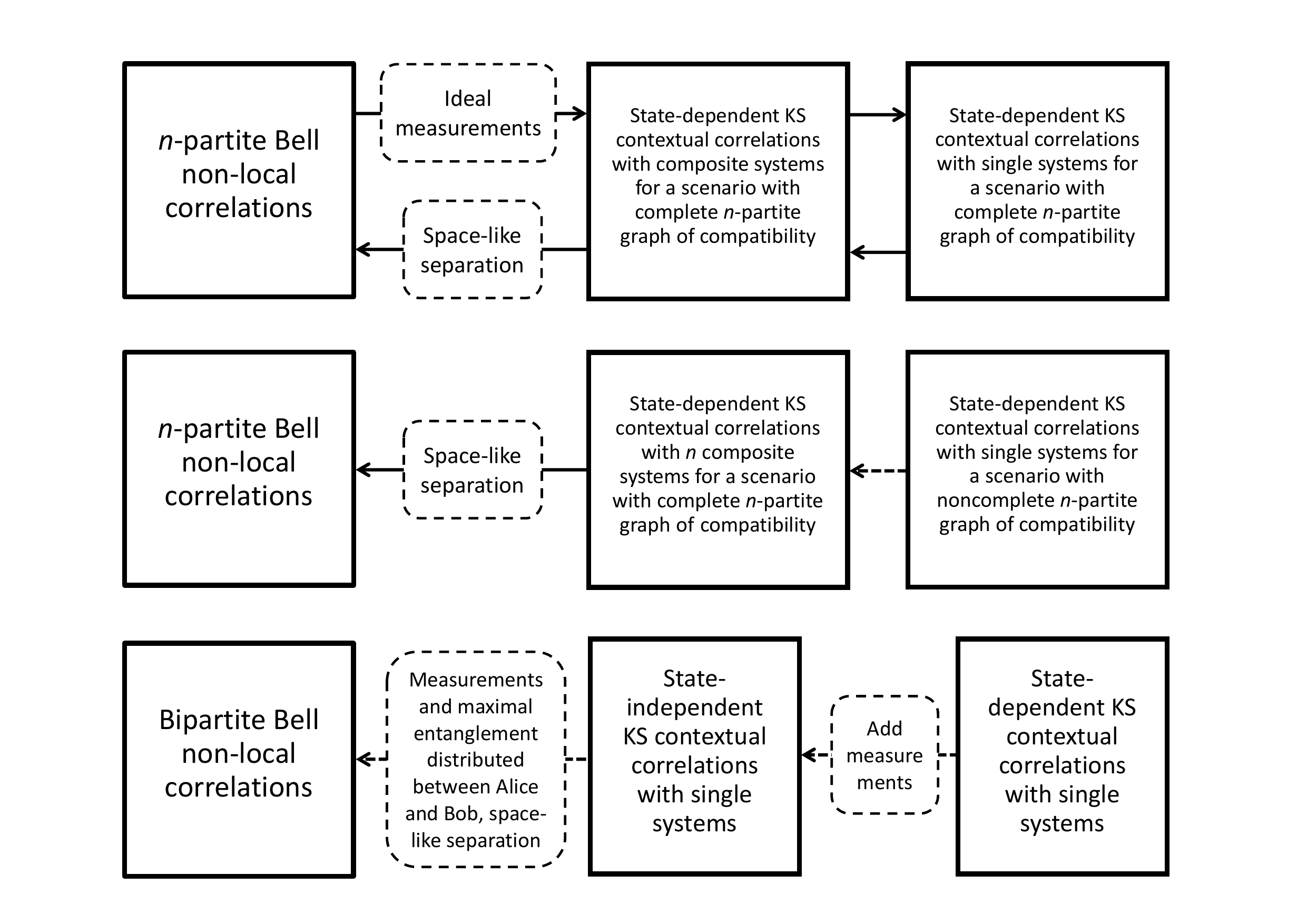}
	\caption{Connections between Bell non-locality and KS contextuality in quantum theory. Continuous (dashed) arrows indicate that the quantum value and the tightness of the inequalities are (not necessarily) preserved. The upper block is discussed in Sects.~\ref{Sec:2} (from Bell non-locality to KS contextuality) and \ref{Sec:3.0} (from KS contextuality to Bell non-locality). The middle block is discussed in Sect.~\ref{Sec:3.0b}. The lower block is discussed in Sect.~\ref{Sec:3.1}.}
	\label{Fig2}
\end{figure}


\section{Connections and implications}
\label{Sec:6}


Now we can collect all the connections between Bell non-locality and KS contextuality that we have discussed so far into a map. This map is shown in Fig.~\ref{Fig2}. 

How this map might be useful? There are, at least, three problems which may benefit from this map.

(A) Classically simulating a particular violation of a Bell inequality requires a certain amount of superluminal communication $C$ \cite{Maudlin92,BCT99,TB03}. Classically simulating a particular violation of a KS non-contextuality inequality has a memory cost memory $M$ \cite{KGPLC11,CGGX18,Budroni19}. The map may help us to formulate the question of what is the relation between $C$ and $M$ in a more precise way, as now, e.g., we can compute $C$ for the violation of the Bell inequality and then $M$ for the corresponding violation of the KS non-contextuality inequality.

(B) In Ref.~\cite{ATC15}, it is shown that ``quantum theory allows for absolute maximal contextuality.'' This means the following. Any KS contextuality witness can be expressed as a sum $S$ of $n$ probabilities of events. The relations of mutual exclusivity between these events can be represented by an $n$-vertex graph $G$ in which there is an edge if the corresponding events are mutually exclusive. The independence number $\alpha(G)$ and the Lov\'asz number $\vartheta(G)$ of $G$ give the maximum of $S$ for non-contextual theories and for quantum theory, respectively \cite{CSW14}. A theory allows for absolute maximal contextuality if it allows that $\vartheta(G) / \alpha(G)$ approach~$n$. The map leads to the following problem: What happens when we extend a quantum absolute maximal contextuality into its minimal state-independent version and then into the corresponding Bell non-locality? How is Bell non-locality in that case? What happens to other interesting forms of contextuality (according to that or other measures of contextuality \cite{GHHHJKW14})?

(C) Self-testing unknown quantum states and measurements is a fundamental problem in quantum information processing. Recently, a method for self-testing using the KCBS KS non-contextuality inequality and its generalizations has been introduced \cite{BRVWCK18}. One practical drawback of the method is the need of assuming that measurements are ideal and perfectly compatible. However, the map guide us to convert the maximal violation of the KCBS KS non-contextuality inequality into a violation of a bipartite Bell inequality. It would be interesting to identify such Bell non-local behaviors and study the connection between self-testing based on KS contextuality \cite{BRVWCK18} and self-testing based on Bell non-locality.

Let this short list serve as an example of the usefulness of the map we have presented. Probably many other problems can take advantage of it.


\section{Acknowledgments}


The author thanks Matthew Pusey and Zhen-Peng Xu for comments on an earlier version of this work. This work was supported by Project Qdisc (Project No.\ US-15097), with FEDER funds, MINECO Project No.\ FIS2017-89609-P, with FEDER funds, and QuantERA grant SECRET, by MINECO (Project No.\ PCI2019-111885-2).




\end{document}